\documentclass[aps,prd,twocolumn,superscriptaddress]{revtex4}
\usepackage{epsfig,epsf}
\usepackage{amsmath}
\usepackage{amsthm}
\usepackage{amsfonts}
\usepackage{amssymb}
\usepackage{dsfont}
\usepackage{multirow}
\usepackage{appendix}
\usepackage{slashed}
\usepackage[active]{srcltx}
\usepackage{psfrag}

\begin{document}

\title{On the nature of $\Xi_c(2930)$}
\date{\today}
\author{T.~M.~Aliev}
\affiliation{ Physics Department, Middle East Technical
University, 06531 Ankara, Turkey}
\author{K.~Azizi}
\affiliation{School of Physics, Institute for Research in
Fundamental Sciences (IPM), P. O. Box 19395-5531, Tehran, Iran}
\affiliation{Department of Physics, Do\v{g}u\c{s} University,
Acibadem-Kadik\"{o}y, 34722 Istanbul, Turkey}
\author{H.~Sundu}
\affiliation{Department of Physics, Kocaeli University, 41380 Izmit, Turkey}

\begin{abstract}
The single charmed excited $\Xi_c(2930)$ state was discovered many years ago by BABAR collaboration and recently confirmed by Belle experiment. However, both of these experiments, unfortunately, could not fix the quantum numbers of this particle and its nature is under debates. In the present study, we calculate its mass and width of its dominant decay to 
$ \Lambda_c K $.  To this end we consider $\Xi_c(2930)$ state once as angularly  excited  and then radially excited single charmed baryon in 
$ \Xi_c$ channel.
 Comparison of the obtained results with the experimental data  suggests assignment of  $\Xi_c(2930)$ state  as the angular excitation of the   ground state $\Xi_c$ baryon with quantum numbers $J^P=\frac{1}{2}^{-}$.
 
 %, however,  to make a final decision on the nature of this state more new experimental data with small statistical %and systematical errors are needed.
%
%The mass and pole residue of the  first orbitally and radially
%excited $ \Xi_c$ state as well as ground state residue are
%calculated by means of the two-point QCD sum rules. Using these
%results the strong coupling constant of the decays
%$\widetilde{\Xi}_c\rightarrow \Lambda_cK$ and $\Xi^{\prime}_c
%\rightarrow \Lambda_c K$ are calculated within light-cone QCD sum
%rules and width of these decay channels are estimated. Obtained
%result for the mass and width of the $\widetilde{\Xi}_c$ baryon is
%in nice consistency with the experimental results obtained by
%Belle Collaboration. From the results of decay widths we conclude
%that the $\Xi_c(2930)$ most probably have negative parity.
\end{abstract}

\maketitle

\section{Introduction}

With impressive developments of experimental techniques many new
conventional and exotic states have been discovered \cite{PDG}. These
discoveries have opened a new direction in hadron physics. Indeed, the heavy
baryon spectroscopy receives special attention as heavy
baryons represent a very suitable place for testing the ground of
heavy quark symmetry and provide us with  deep
understanding of the details of the strong interaction.

While many new excited charmed baryons are discovered in  different experiments and many
theoretical works are devoted to establishing their quantum numbers, still their nature is not evident, and 
many open questions remain on their internal structure and quark organization. For instance, there have been made many suggestions on the structures of the newly observed five narrow resonances $\Omega_c(3000), \Omega_c(3050), \Omega_c(3066), \Omega_c(3090), \Omega_c(3119)$  by LHCb Collaboration in the $\Xi_c^+ K^-$ invariant mass spectrum \cite{Aaij:2017nav}: some authors have treated them as usual three-quark resonances \cite{Agaev:2017jyt,Agaev:2017lip,Aliev:2017led,Karliner:2017kfm,Wang:2017vnc}, while some others have interpreted them as  new penta-quark states \cite{Yang:2017rpg,Huang:2017dwn}.
At present, there is unfortunately no any phenomenological model, which can
successfully describe the properties of such complicated systems
\cite{Crede:2013sze,Cheng}. For this,  more experimental and
theoretical attempts are needed to understand dynamics of these
new systems.

Some new excited states at $ \Xi_c $, $ \Sigma_c $ and $ \Lambda_c $ channels have also been discovered that are  of great importance and deserve  investigations with the aim of clarification of their nature and internal structure. 
The observation of the charmed-strange baryon $\Xi_c(2930)$, which is the subject of the present study, has a long history in the experiment. This state  was firstly observed by BABAR
Collaboration in 2008 with mass
$m=[2931\pm3(\mathrm{stat.})\pm5(\mathrm{syst.})]~~\mathrm{MeV}/c^2$ and width of
$\Gamma=[36\pm7(\mathrm{stat.})\pm11(\mathrm{syst.})]~\mathrm{MeV}$ as an intermediate resonance in the decay  $B^-\rightarrow
\Lambda_c^+\bar{\Lambda}_c^-K^-$ \cite{Aubert}. Note that the Belle Collaboration had before measured the branching ratios of the decays  $B^+\rightarrow
\Lambda_c^+\Lambda_c^-K^+$ and $B^0\rightarrow
\Lambda_c^+\Lambda_c^-K^0$ in 2006 \cite{Gabyshev} but could not find any intermediate charmed resonances. After  observation of $\Xi_c(2930)$ by BABAR, the state $\Xi_c(2930)$ was investigated in the framework of different theoretical models like constituent quark model
\cite{Wang:2017kfr,Chen:2016iyi}, chiral quark model \cite{Liu},
QCD sum rules \cite{Chen}, etc.

%Next, this state looking for in analysis of $B^-\rightarrow
%K^-\Lambda_c^+\bar{\Lambda}_c^-$ decay by Belle Collaboration
%\cite{Gabyshev}, but it was not observed. 
Very recently, Belle
Collaboration performed an updated measurement on $B^-\rightarrow
K^-\Lambda_c^+\bar{\Lambda}_c^-$decay   and observed the $\Xi_c(2930)$
state in the $ K^-\Lambda_c^+ $ invariant mass with a significance of $5.1\sigma$ \cite{Li:2017uvv}. The
measured mass and width is:
\begin{eqnarray}
m&=&[2928.9\pm3.0(\mathrm{stat.})^{+0.8}_{-12.0}(\mathrm{syst.})]\mathrm{MeV/c^2}
\nonumber \\
\Gamma&=&[19.5\pm8.4(\mathrm{stat.})^{+5.4}_{-7.9}(\mathrm{syst.})]\mathrm{MeV},
\end{eqnarray}
respectively. However, both of the experiments could not, unfortunately, fix the quantum numbers of $\Xi_c(2930)$ state.  This automatically suggests more experimental and  theoretical
efforts on the properties of this resonance.

We aim to calculate some parameters of $\Xi_c(2930)$ state in the present study to clarify its nature and fix its quantum numbers. To this end, we assume it once as angularly excited  negative parity  ($ \widetilde{\Xi}_c  $) and the second as radially excited    positive parity  ($ \Xi_c^{\prime} $) spin-1/2 baryon at  $ \Xi_c $ channel.  In quark model's notations these states are represented by $1^2 P_{1/2}$ and $2^2 S_{1/2}$,
respectively. For customary, in next discussions, we will denote these states  as $J^P=1/2^- $ and $J^P=1/2^+$, respectively.  We evaluate the widths of the strong decays $
\widetilde{\Xi}_c \rightarrow \Lambda_c K$ and
$\Xi_c^{\prime}\rightarrow \Lambda_cK$. For this, 
firstly we compute the mass and residue of the ground state, first
angularly and radially excited $\Xi_c$ baryons as well as the
couplings of the strong  $ \widetilde{\Xi}_c \Lambda_c K$ and
$\Xi_c^{\prime}\Lambda_cK$ vertices allowing us to find the
required decay widths.  For calculation of the masses and residues we
employ QCD two-point sum rule, whereas in the case of the strong
couplings we apply the technique of QCD light-cone sum rule (LCSR). Note that using a $ ^3P_0 $ model the authors in \cite{Ye:2017dra,Ye:2017yvl} have concluded that the resonance $\Xi_c(2930)$ may be P-wave, D-wave or 2S-wave
excitation of  the ground state $\Xi_c$ baryon with different quantum numbers, $ J^P=\frac{1}{2}^{\pm} $, $ \frac{3}{2}^{\pm} $ or $ \frac{5}{2}^{-} $  by analyses of different excitations of charmed strange baryons. In order to distinguish among these possibilities, they have suggested measurements of some ratios of the branching fractions  associated to some possible decay modes of the $\Xi_c(2930)$ state.

The article is organized in the following way. In section II, the
mass sum rules for $\Xi_c$ baryons including its first angular and
radial excitations are calculated, and the values of the masses and residues are found. Section III is devoted to
the calculation of the strong coupling constants defining the $ \widetilde{\Xi}_c \Lambda_c K$ and
$\Xi_c^{\prime}\Lambda_cK$ vertices. We estimate
the widths of the decay channels under consideration and compare the results obtained on the masses and widths with the experimental data with the aim of fixing the quantum numbers of the $\Xi_c(2930)$ resonance.
The last section is reserved for  summary and concluding remarks.

%%%%%%%%%%%%%%%%%%%%%%%%%%%%%%%%%%%%

\section{Masses and Pole Residues of the  first angularly and radially excited $\Xi_c$ states}

As we noted,  the $\Xi_c(2930)$ has been seen as a peak in the
$\Lambda_c^+K^-$ invariant mass distribution. But unfortunately,
its quantum numbers  have not established yet. In
present work, we consider two possible scenarios for it: a) The
$\Xi_c(2930)$ is considered as the radial excitation of the ground state
$\Xi_c(2467)$. In other words it carries the same quantum numbers as
$\Xi_c(2467)$, i.e. $J^P=\frac{1}{2}^+$. b) The $\Xi_c(2930)$ is
treated as the first angular excitation of the $\Xi_c(2467)$, that is 
negative parity baryon with $J^P=\frac{1}{2}^-$. Note that, in the following,   we will consider in more details  the second scenario. The  results for  the first scenario will be obtained by some replacements that will be mentioned later. Here we should also note that in \cite{Mao:2015gya} the P-wave heavy baryon masses are calculated with QCD sum rules in the framework of the heavy quark effective theory. 

In order to calculate the mass and residue of  $\Xi_c$ baryon, we
start with the following two point correlation function:
\begin{equation}
\Pi (q)=i\int d^{4}xe^{iq\cdot x}\langle 0|\mathcal{T}\Big\{\eta_{\Xi_c }(x)%
\bar{\eta}_{\Xi_c }(0)\Big\}|0\rangle ,  \label{eq:CorrF1}
\end{equation}%
where $\eta_{\Xi_c }(x)$ is the interpolating current for $\Xi_c $
state with spin-parity $J^P=\frac{1}{2}^+$ and $\mathcal{T}$ indicates
the time ordering operator. The general form of the interpolating
current for the heavy spin-${1\over 2}$, $\Xi_c$ baryon belonging
to antitriplet representations of $SU(3)$ can be written as:

\begin{eqnarray}
 \label{Eq:Current}
\eta_{\Xi_c} &=& {\epsilon^{abc}\over \sqrt{6}} \Big\{ 2 \Big(
q_1^{aT} C q_2^b \Big) \gamma_5 c^c + 2 \beta \Big( q_1^{aT} C
\gamma_5 q_2^b \Big) c^c
 \nonumber \\
&+& \Big( q_1^{aT} C c^b \Big) \gamma_5 q_2^c + \beta
\Big(q_1^{aT} C
\gamma_5 c^b \Big) q_2^c \nonumber \\
&+& \Big(c^{aT} C q_2^b \Big) \gamma_5 q_1^c + \beta \Big(c^{aT} C
\gamma_5 q_2^b \Big) q_1^c \Big\}~,
\end{eqnarray}
 where $a,,b,c$
are the color indices, $ C $ is the charge conjugation operator and  $\beta$ is an arbitrary parameter with
$\beta =-1$ corresponding to the Ioffe current. $q_1$ and $q_2$
are $u(d)$ and $s$ quarks for $\Xi_{c}^{+}(\Xi_{c}^{0})$ baryon,
respectively. Here some details about the above current are in order. According to the quark model, $ \Xi_{c} $ belongs to the antitriplet representation of the $SU(3)$, i.e. the current describing this state should be antisymmetric with respect to the exchange of the light quarks' fields. The interpolating current must also be a color singlet. Therefore, its general form satisfying both these conditions can be written as

\begin{eqnarray}
\label{kazem}
\eta_{\Xi_c} &\sim & \epsilon^{abc} \Big\{  \Big(
q_1^{aT} C \Gamma q_2^b \Big) \tilde{\Gamma} c^c 
+ \Big( q_1^{aT} C \Gamma c^b \Big) \tilde{\Gamma} q_2^c\nonumber\\ 
&-& \Big(q_2^{aT} C\Gamma c^b \Big) \tilde{\Gamma} q_1^c  \Big\}~,
\end{eqnarray}
where $ \Gamma$, $ \tilde{\Gamma}=$ $ 1 $, $ \gamma_5 $, $ \gamma_\mu $, $ \gamma_5 \gamma_\mu$ or $ \sigma_{\mu\nu} $. We need to determine 
$ \Gamma$ and $ \tilde{\Gamma}$. To this end let us first consider the  transpose of the  term $ \epsilon^{abc} (q_1^{aT} C \Gamma q_2^b )  $:
\begin{eqnarray}
  [\epsilon^{abc} q_1^{aT} C \Gamma q_2^b  ]^T=\epsilon^{abc} q_2^{bT} C (C \Gamma C^{-1}) q_1^a,
\end{eqnarray}
where $ C \Gamma C^{-1} $ is equal to $ \Gamma $ for $\Gamma =1 $, $ \gamma_5 $ or $ \gamma_5 \gamma_\mu$ and it is equal to 
$ -\Gamma $ for $ \Gamma=\gamma_\mu $  or $ \sigma_{\mu\nu} $.  The transpose of a one by one matrix should be equal to itself, i.e.
\begin{eqnarray}
  \epsilon^{abc} q_1^{aT} C \Gamma q_2^b  =-\epsilon^{abc} q_2^{aT} C  \Gamma  q_1^b,
\end{eqnarray}
for  $\Gamma =1 $, $ \gamma_5 $ or $ \gamma_5 \gamma_\mu$. In result, indeed $ \epsilon^{abc} q_1^{aT} C \Gamma q_2^b  $ is antisymmetric for the $ q_1\leftrightarrow q_2 $   replacement  if $\Gamma =1 $, $ \gamma_5 $ or $ \gamma_5 \gamma_\mu$.

The simplest way is to take the  $ \Xi_{c} (2930) $ to have the same total spin and spin projection as the charm quark. Thus the spin of the diquark formed by light quarks is zero. This implies $\Gamma =1 $ or $ \gamma_5 $. Therefore, the two possible forms of the interpolating current can be written as:
\begin{eqnarray}
\eta_{1} &= & \epsilon^{abc}   \Big(
q_1^{aT} C  q_2^b \Big) \tilde{\Gamma}_1 c^c 
\nonumber\\ 
\eta_{2} &= & \epsilon^{abc}   \Big(
q_1^{aT} C\gamma_5  q_2^b \Big) \tilde{\Gamma}_2 c^c.
\end{eqnarray}
The forms of $ \tilde{\Gamma}_1 $ and $  \tilde{\Gamma}_2$ are determined through the Lorentz and parity considerations. Since $ \eta_{1} $ and $ \eta_{2} $ are Lorentz scalars, one must have $ \tilde{\Gamma}_1 $, $  \tilde{\Gamma}_2=$  $1 $ or $ \gamma_5 $. The parity transformation leads to the result that $ \tilde{\Gamma}_1 =\gamma_5$ and $ \tilde{\Gamma}_2 =1$. Therefore the two possible forms of the interpolating current are
\begin{eqnarray}
\eta_{1} &= & \epsilon^{abc}   \Big(
q_1^{aT} C  q_2^b \Big) \gamma_5 c^c 
\nonumber\\ 
\eta_{2} &= & \epsilon^{abc}   \Big(
q_1^{aT} C\gamma_5  q_2^b \Big)  c^c.
\end{eqnarray}
Obviously their arbitrary linear combination can better represent the baryon under consideration, i.e.
\begin{eqnarray}
\eta &\sim & \epsilon^{abc} \Big[  \Big(
q_1^{aT} C  q_2^b \Big) \gamma_5 c^c 
+ \beta   \Big(
q_1^{aT} C\gamma_5  q_2^b \Big)  c^c\Big],
\end{eqnarray}
where we introduced the general parameter $ \beta $ to obtain the most general form of $ \eta $. Performing similar analyses for the second and third terms in Eq. (\ref{kazem}) and with the combinations presented in Eq. (\ref{Eq:Current}) we get the most general form of the interpolating current for $ \Xi_{c} (2930) $.

To derive the mass sum rules for the $\Xi_c(2930) $ baryon we
calculate the correlation function using two languages: 
 hadronic, in terms of the masses and residues called the physical side and QCD, in terms of the fundamental QCD degrees of freedom called the QCD or theoretical side. By equating these two representations,
one can get the QCD sum rules for the physical quantities of the
baryons under consideration.  The physical
side of the correlation function is obtained by inserting the
complete sets of intermediate states with  both parities:
\begin{eqnarray}
\Pi ^{\mathrm{Phys}}(q) &=&\frac{\langle 0|\eta_{\Xi_c}|\Xi_c (q,s)\rangle
\langle
\Xi_c (q,s)|\overline{\eta}_{\Xi_c}\rangle }{m^{2}-q^{2}}  \notag \\
&+&\frac{\langle 0|\eta_{\Xi_c}|\widetilde{\Xi }_c(q,\widetilde{s})\rangle
\langle
\widetilde{\Xi }_c(q,\widetilde{s})|\overline{\eta}_{\Xi_c}\rangle }{\widetilde{m}%
^{2}-q^{2}}  \notag \\
&+&\ldots ,  \label{eq:CF1/2}
\end{eqnarray}%
where $m$, $\widetilde{m}$ and $s$, $\widetilde{s}$ are the masses
and spins of the ground and first angularly excited $\Xi_c $
baryons, respectively. The dots denote contributions of higher
resonances and continuum states. In Eq.\ (\ref{eq:CF1/2}) the
summations over the spins $s$ are $\widetilde{s}$ are implied.

The matrix elements in Eq. (\ref{eq:CF1/2}) are determined as
\begin{eqnarray}
\langle 0|\eta_{\Xi_c}|\Xi_c (q,s)\rangle  &=&\lambda u(q,s),  \notag \\
\langle 0|\eta_{\Xi_c}|\widetilde{\Xi }_c(q,\widetilde{s})\rangle  &=&\widetilde{%
\lambda }\gamma _{5}u^{-}(q,\widetilde{s}).  \label{eq:MElem}
\end{eqnarray}%
Here $\lambda $ and $\widetilde{\lambda }$ are the residues of the
ground and first angularly excited $\Xi_c $ baryons, respectively.
Using Eqs.\ (\ref {eq:CF1/2}) and (\ref{eq:MElem}) and carrying
out summations over the spins of corresponding baryons,
we obtain
\begin{equation}
\Pi ^{\mathrm{Phys}}(q)=\frac{\lambda^2 (\slashed q+m)}{m^{2}-q^{2}}+\frac{%
\widetilde{\lambda }^2(\slashed q-\widetilde{m})}{\widetilde{m}^{2}-q^{2}}%
+\ldots.
\end{equation}
Performing  Borel transformation of this expression we have
\begin{eqnarray}
\mathcal{B}\Pi ^{\mathrm{Phys}}(q) &=&\lambda ^{2}e^{-\frac{m^{2}}{M^{2}}}(%
\slashed q+m)  \notag \\
&+&\widetilde{\lambda }^2e^{-\frac{\widetilde{m}^{2}}{M^{2}}}(\slashed q-%
\widetilde{m}).  \label{eq:Bor1}
\end{eqnarray}

The QCD side of the aforementioned correlation function is
calculated in terms of the QCD degrees of freedom in deep
Euclidean region. After
inserting the explicit form of the interpolating current given by Eq. (\ref%
{Eq:Current}) into the correlation function in Eq.
(\ref{eq:CorrF1}) and performing contractions via the Wick's
theorem, we get the QCD side in terms of the light and heavy
quarks propagators. By using light and heavy quark propagators in
the coordinate space and performing the Fourier and Borel
transformations, as well as applying the continuum subtraction,
after  lengthy calculations for the correlation function we obtain
\begin{equation}
\mathcal{B}\Pi ^{\mathrm{QCD}}(q)=\mathcal{B}\Pi
_{1}^{\mathrm{QCD}} \slashed q+\mathcal{B}\Pi
_{2}^{\mathrm{QCD}}I.
\end{equation}
The  expressions  for  $\mathcal{B}\Pi _{1}^{\mathrm{QCD}}$ and
$\mathcal{B}\Pi _{2}^{\mathrm{QCD}}$ are presented in Appendix.

Having calculated both the hadronic and QCD sides of the
correlation function, we match the coefficients of the structures
$\slashed q$ and $I$ from these two sides and obtain the following
sum rules, which are used to extract the masses and residues of the
ground and first angularly excited states:
\begin{eqnarray}
\lambda^{2}e^{-\frac{m^{2}}{M^{2}}}+\widetilde{\lambda}^{2}e^{-\frac{
\widetilde{m}^{2}}{M^{2}}}&=&\mathcal{B}\Pi _{1}^{\mathrm{QCD}},
\notag
\\
\lambda
^{2}me^{-\frac{m^{2}}{M^{2}}}-\widetilde{\lambda}^{2}\widetilde{m}e^{-\frac{
\widetilde{m}^{2}}{M^{2}}}&=&\mathcal{B}\Pi _{2}^{\mathrm{QCD}}.
\label{eq:MFor1}
\end{eqnarray}
Using  two equations given in Eq. (\ref{eq:MFor1})
 it is  easy  to show that
\begin{eqnarray}
\widetilde{m}^{2}&=&\frac{\mathcal{B}\widetilde{\Pi}
_{2}^{\mathrm{QCD}}-m\mathcal{B}\widetilde{\Pi}
_{1}^{\mathrm{QCD}}}{\mathcal{B}\Pi
_{2}^{\mathrm{QCD}}-m\mathcal{B}\Pi _{1}^{\mathrm{QCD}}},
\nonumber \\
\widetilde{\lambda}^{2}&=&\frac{m\mathcal{B}\Pi
_{1}^{\mathrm{QCD}}-\mathcal{B}\Pi
_{2}^{\mathrm{QCD}}}{m+\widetilde{m}}e^{\frac{\widetilde{m}^2}{M^2}},
\nonumber \\
\lambda^2&=&\frac{\widetilde{m}\mathcal{B}\Pi
_{1}^{\mathrm{QCD}}+\mathcal{B}\Pi
_{2}^{\mathrm{QCD}}}{m+\widetilde{m}}e^{\frac{m^2}{M^2}},
 \label{Eq:massResidue}
\end{eqnarray}
where $\mathcal{B}\widetilde{\Pi}
_{1(2)}^{\mathrm{QCD}}=\frac{d\mathcal{B}\Pi
_{1(2)}^{\mathrm{QCD}}}{d(-\frac{1}{M^2})}$.

For obtaining the expressions for the mass and residue for
radially excitation state it is enough to make replacement
$\widetilde{m}\rightarrow -m^{\prime}$ and redefine the residue
$\widetilde{\lambda}$ as $\lambda^{\prime}$ in expressions of Eq.
(\ref{Eq:massResidue}).

\begin{table}[tbp]
%\rowcolors{1}{lightgray}{white}
\begin{tabular}{|c|c|}
\hline\hline Parameters & Values \\ \hline\hline
$m_{c}$ & $(1.28\pm0.03)~\mathrm{GeV}$ \cite{PDG}\\
$m_{s}$ & $96^{+8}_{-4}~\mathrm{MeV}$ \cite{PDG}\\
$m_{\Xi_c(2467)}$ & $(2467.87\pm0.30)~\mathrm{MeV}$ \cite{PDG}\\
$\langle \bar{q}q \rangle $ & $(-0.24\pm 0.01)^3$ $\mathrm{GeV}^3$  \\
$\langle \bar{s}s \rangle $ & $0.8\cdot(-0.24\pm 0.01)^3$ $\mathrm{GeV}^3$  \\
$\langle \overline{q}g_s\sigma Gq\rangle$ & $m_{0}^2\langle
\bar{q}q \rangle$
\\
$\langle \overline{s}g_s\sigma Gs\rangle$ & $m_{0}^2\langle
\bar{s}s \rangle$
\\
$m_{0}^2 $ & $(0.8\pm0.1)$ $\mathrm{GeV}^2$ \\
$\langle\frac{\alpha_sG^2}{\pi}\rangle $ & $(0.012\pm0.004)$ $~\mathrm{GeV}%
^4 $\\
\hline\hline
\end{tabular}%
\caption{Some input parameters used in the calculations.}
\label{tab:Param}
\end{table}

\begin{widetext}

\begin{figure}[h!]
\begin{center}
\includegraphics[totalheight=6cm,width=8cm]{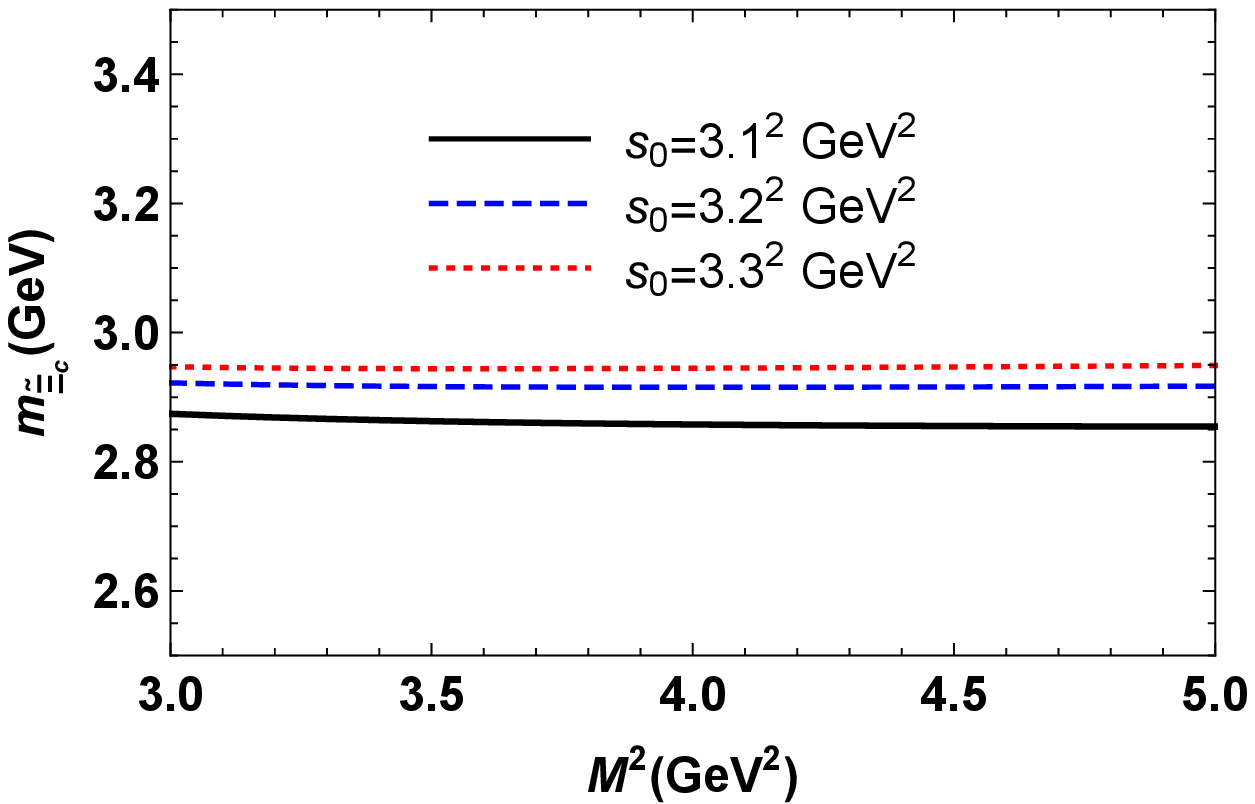}
\includegraphics[totalheight=6cm,width=8cm]{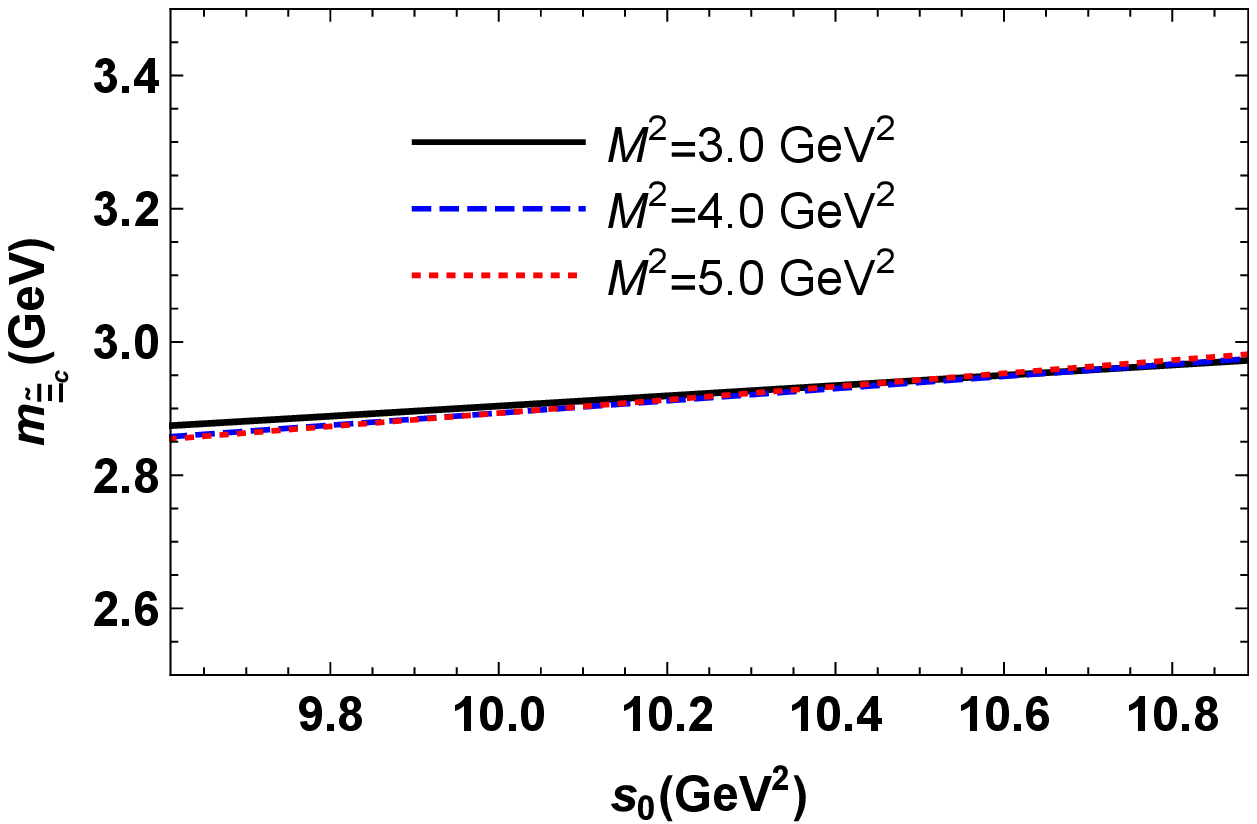}
\end{center}
\caption{ The mass of the $\widetilde{\Xi}_c $ baryon as a
function of the Borel parameter $M^2$ at chosen values of $s_0$
(left panel), and as a function of the continuum threshold $s_0$
at fixed $M^2$ (right panel) with $\beta=-0.75$.}
\label{gr:mass1P}
\end{figure}

\begin{figure}[h!]
\begin{center}
\includegraphics[totalheight=6cm,width=8cm]{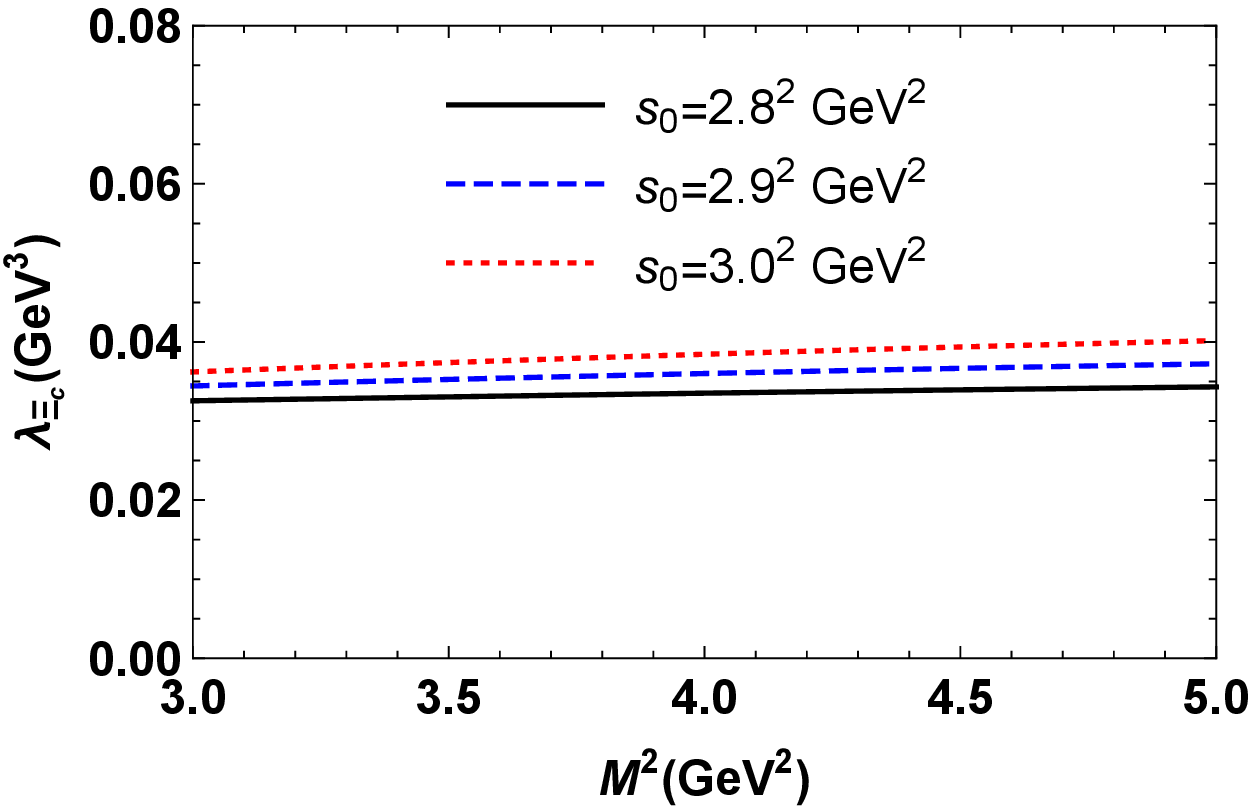}
\includegraphics[totalheight=6cm,width=8cm]{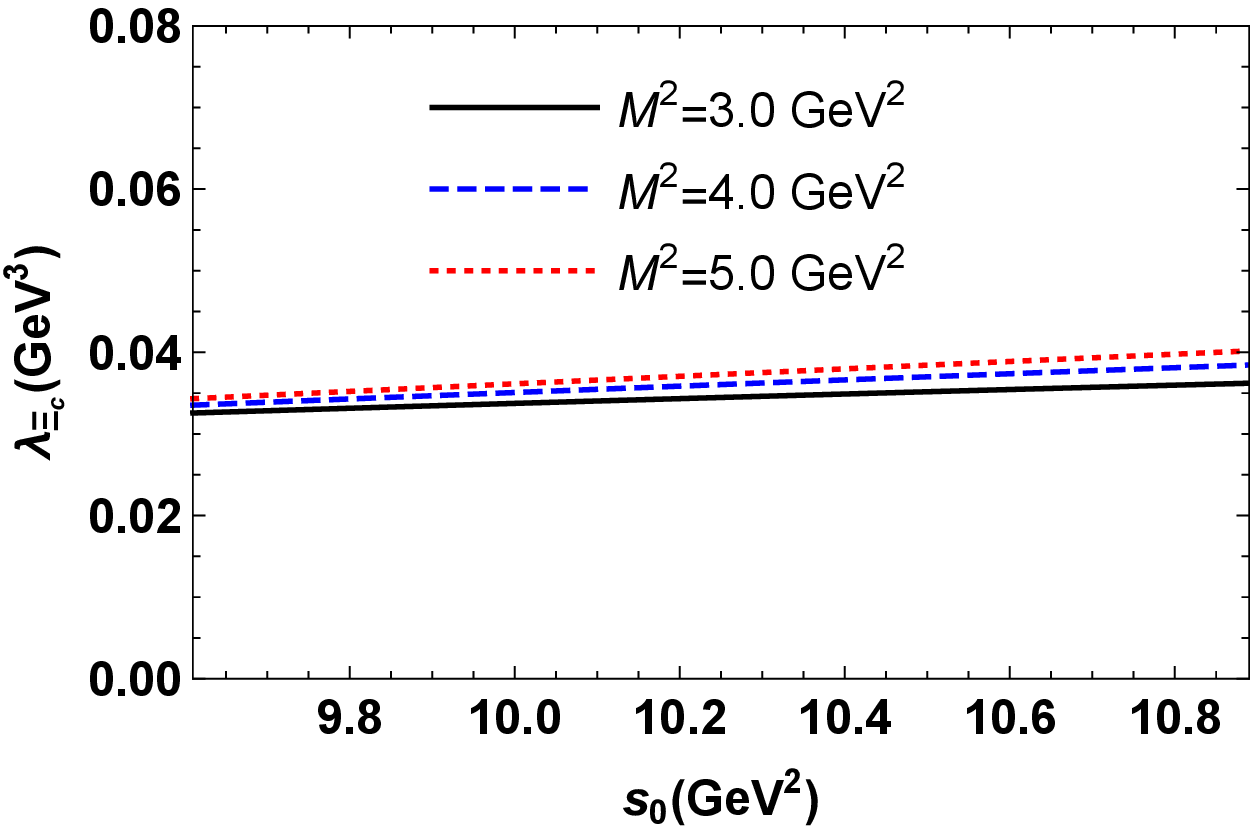}
\end{center}
\caption{ The residue of the $\Xi_c$ baryon as a function of the
Borel parameter $M^2$ at chosen values of $s_0$ (left panel), and
as a function of the continuum threshold $s_0$ at fixed $M^2$
(right panel) with $\beta=-0.75$.} \label{gr:GroundLam}
\end{figure}

\begin{figure}[h!]
\begin{center}
\includegraphics[totalheight=6cm,width=8cm]{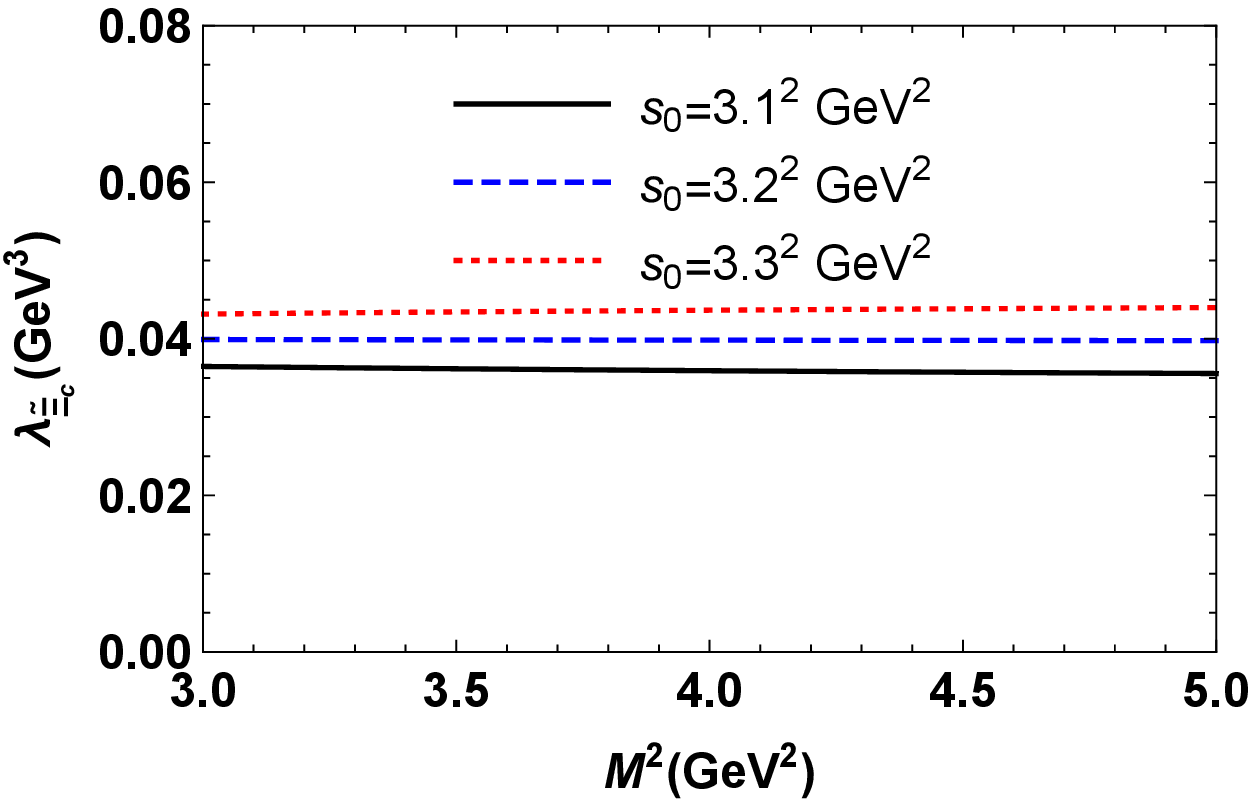}
\includegraphics[totalheight=6cm,width=8cm]{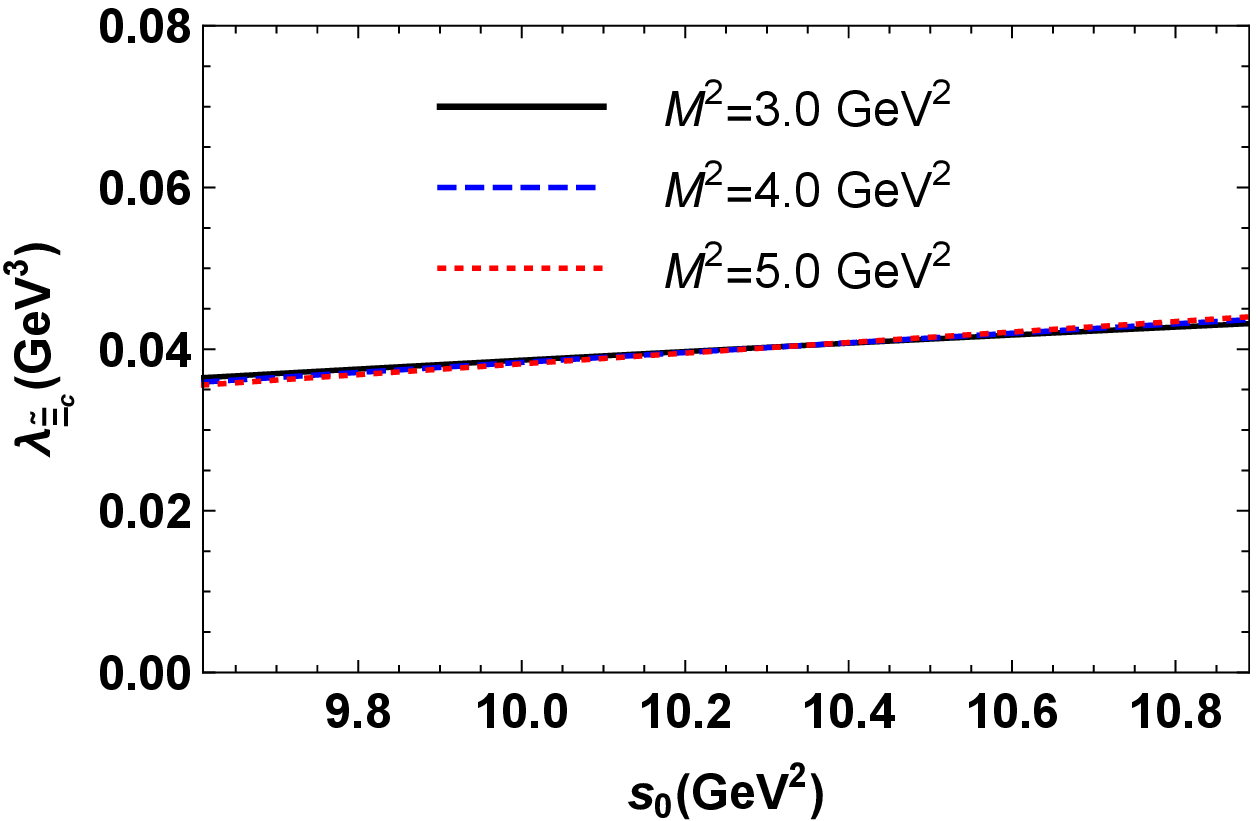}
\end{center}
\caption{The same as in Fig. \ref{gr:GroundLam}, but for the
angularly excited $\widetilde{\Xi}_c$ baryon.}
\label{gr:lam1P}
\end{figure}

\begin{figure}[h!]
\begin{center}
\includegraphics[totalheight=6cm,width=8cm]{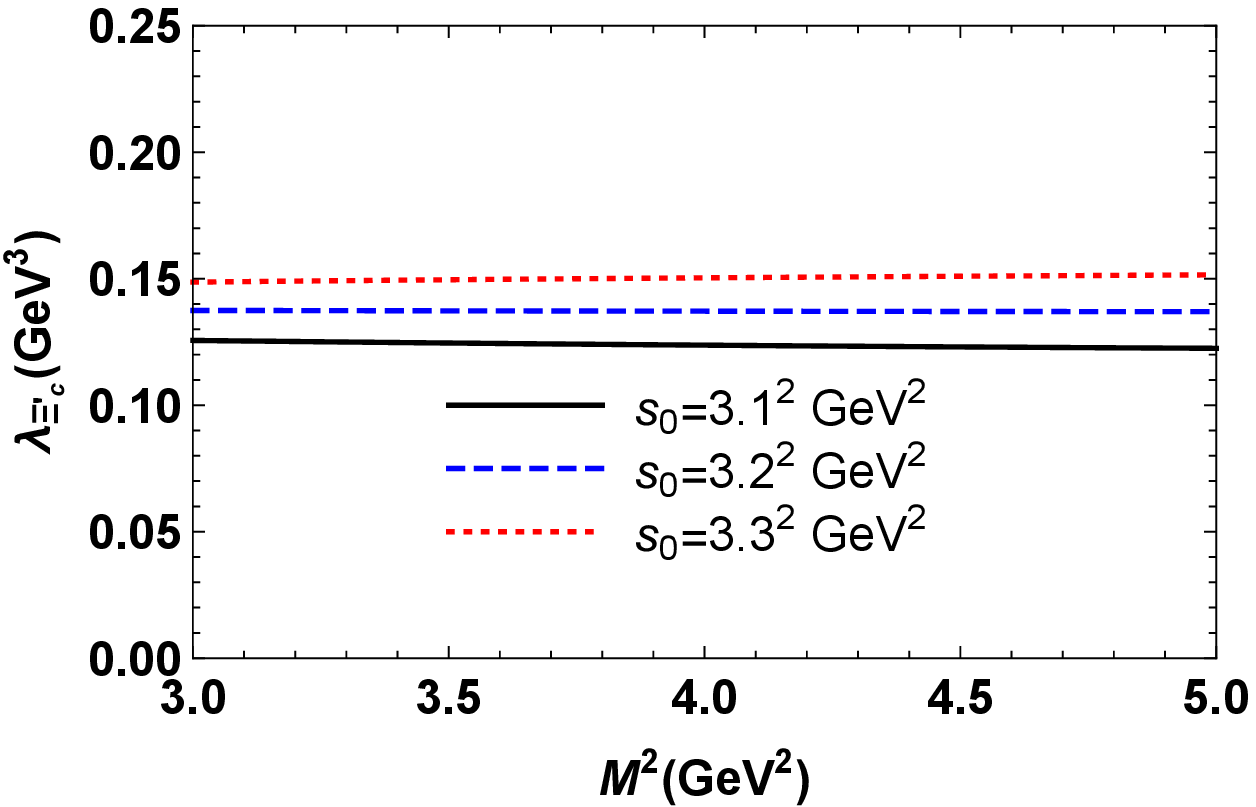}
\includegraphics[totalheight=6cm,width=8cm]{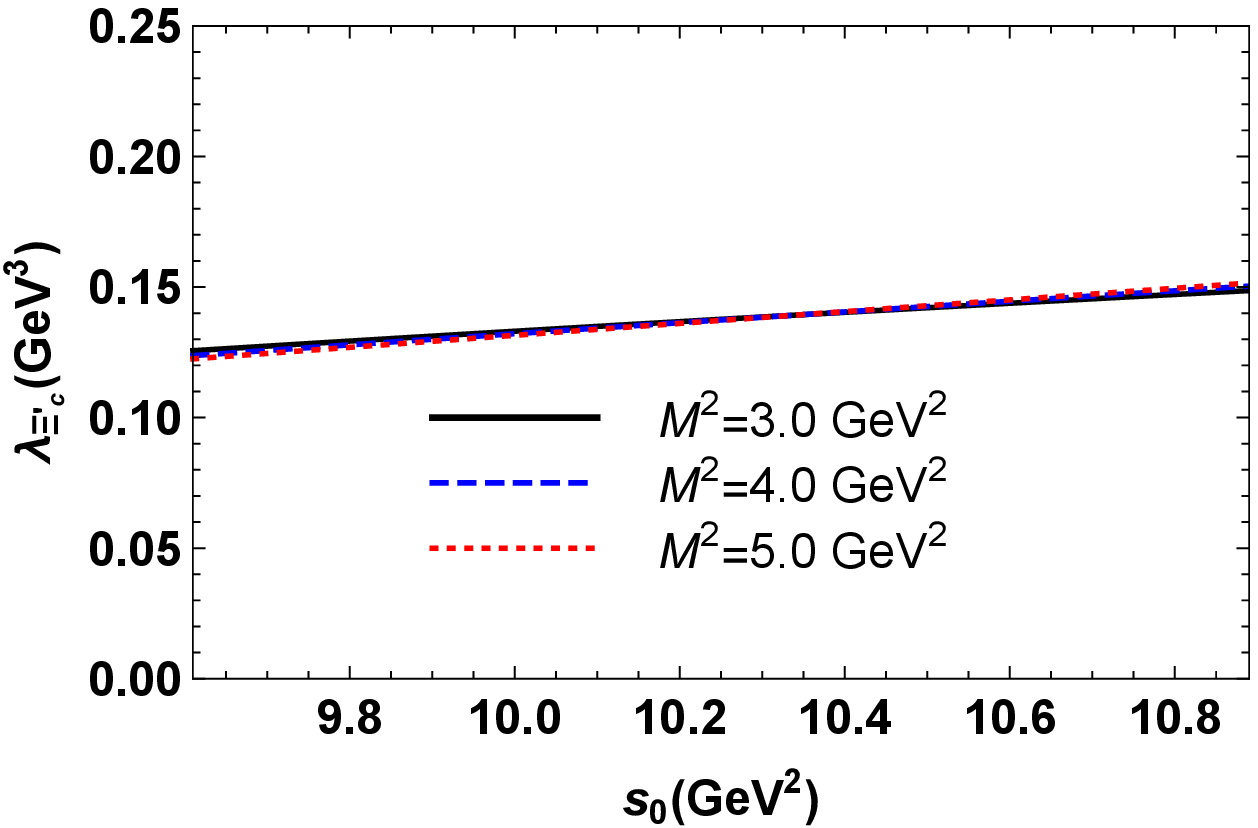}
\end{center}
\caption{The same as in Fig. \ref{gr:GroundLam}, but for the
radially excited $\Xi'_c$ baryon.}
\label{gr:lam2S}
\end{figure}

%\begin{figure}[h!]
%\begin{center}
%\includegraphics[totalheight=6cm,width=8cm]{grlamgroundvsCosteta.eps}
%\end{center}
%\caption{The residue of the ground state $\Xi_c$ baryon as a
%function of  $\cos\theta$ at central values of the $s_0$ and
%$M^2$.} \label{gr:lamGorundCosTeta}
%\end{figure}

\begin{figure}[h!]
\begin{center}
\includegraphics[totalheight=6cm,width=8cm]{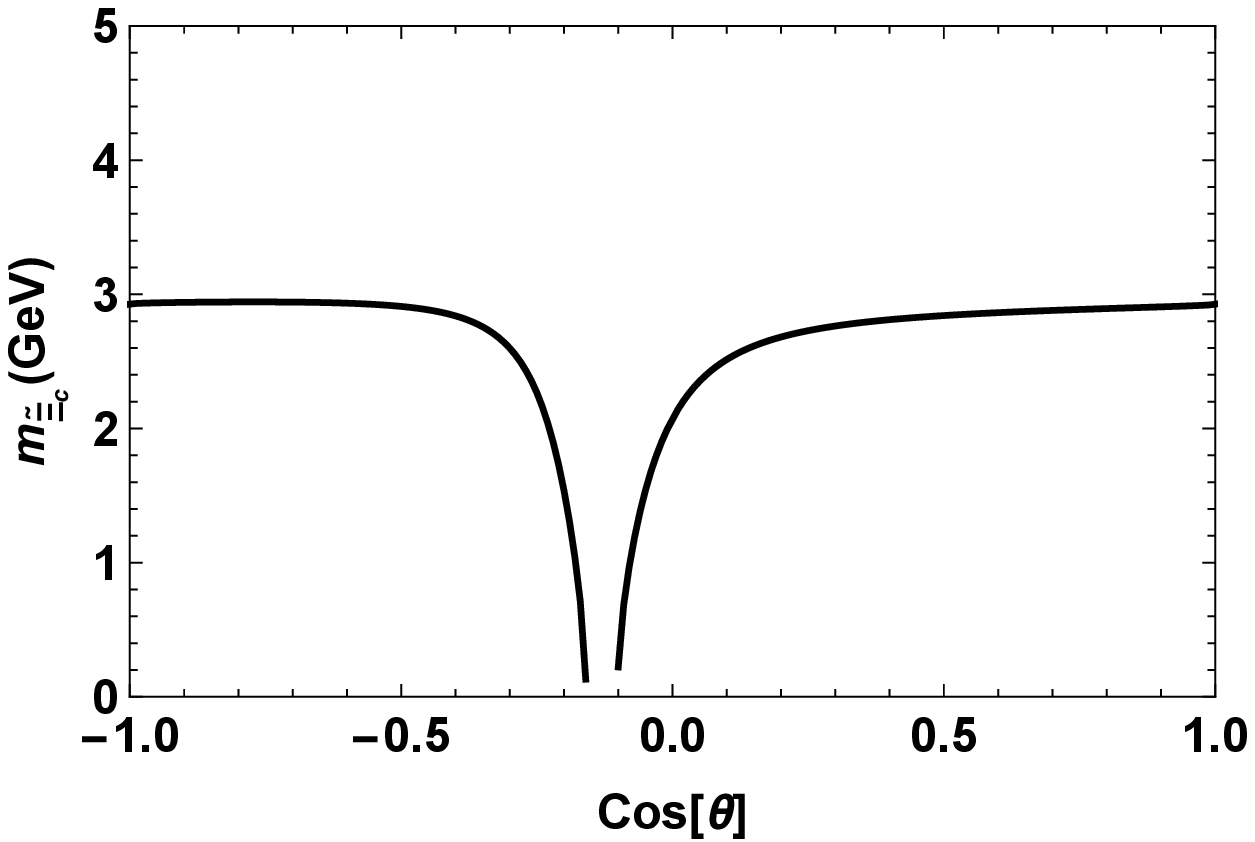}
\includegraphics[totalheight=6cm,width=8cm]{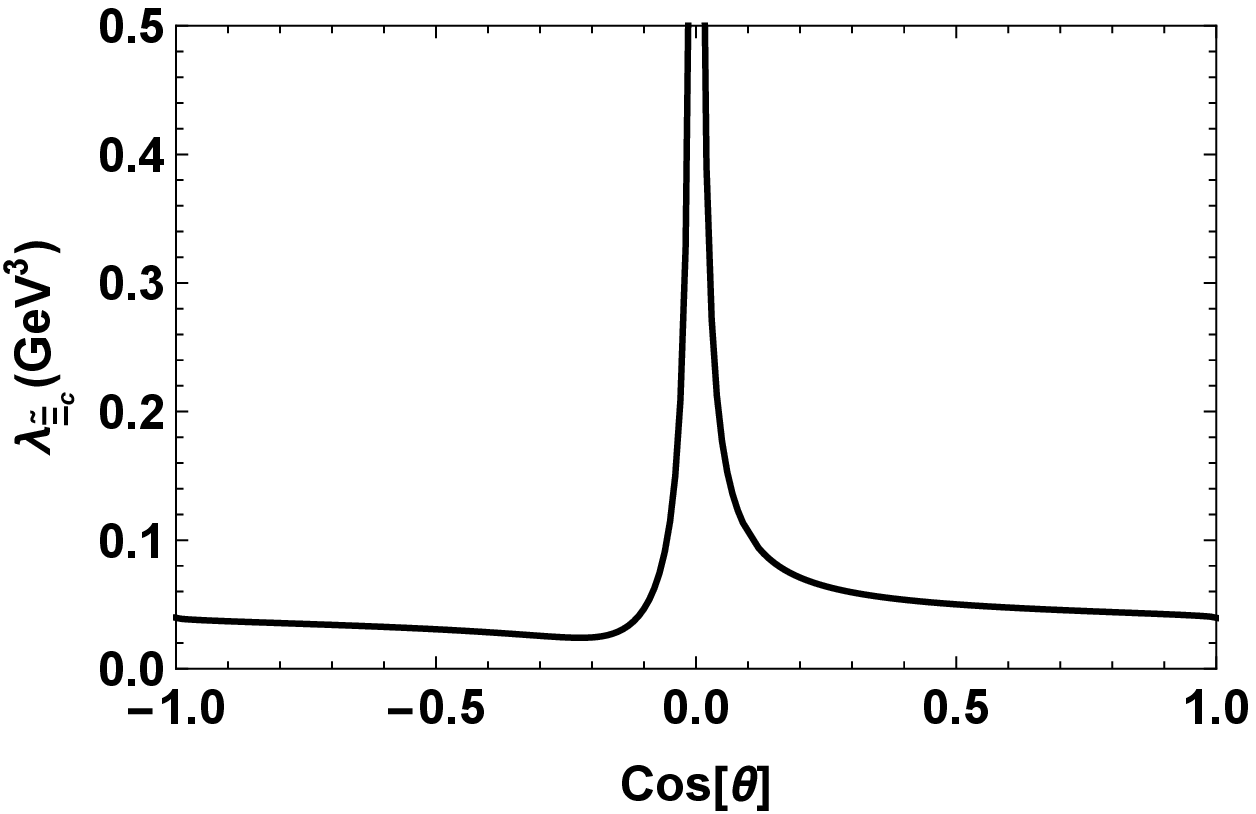}
\end{center}
\caption{  The mass of the $\widetilde{\Xi}_c$ baryon as a
function of the $\cos\theta$ at central values of the $s_0$ and
$M^2$ (left panel). The residue of the $\widetilde{\Xi}_c$ baryon
as a function of  $\cos\theta$ at central values of the $s_0$
and $M^2$ (right panel).} \label{gr:lam1PCosTeta}
\end{figure}

\end{widetext}

To perform analysis of  the sum rules for the masses and residues
of the angularly and radially excited state of the $\Xi_c$ baryon
as well as the residue of the ground state we need some inputs
which are presented in Table~\ref{tab:Param}. The mass of the ground
state $\Xi_c$ is also  taken as an input parameter. Besides the input
parameters, QCD sum rules contains three auxiliary parameters
namely the continuum threshold $s_0$, Borel parameter $M^2$ and
an arbitrary mixing parameter $\beta$. The working windows of these
parameters are determined by demanding that the physical quantities
under consideration are roughly independent of these parameters. To
assess the working interval of the Borel parameter $M^2$ one
needs to consider two criteria: convergence of the series of operator product expansion (OPE)
and adequate suppression of the higher states and continuum.
Consideration of these criteria in the analysis leads to the
following working interval  of $M^2$: 
\begin{eqnarray}
3\ \mathrm{GeV}^{2}\leq M^{2}& \leq &5\ \mathrm{GeV}^{2}.
\label{Eq:MsqMpsq}
\end{eqnarray}%
To determine the working region of the continuum threshold, we
impose the conditions of the pole dominance and OPE  convergence. This
leads to the interval
\begin{eqnarray}
3.1^2\,\,\mathrm{GeV}^{2}&\leq& s_{0}\leq
3.3^2\,\,\mathrm{GeV}^{2}. \label{Eq:s0s0p}
\end{eqnarray}

In order to explore the sensitivity of the obtained results on the
Borel parameter $M^2$ and continuum threshold $s_0$, as examples, in Figs.
\ref{gr:mass1P}-\ref{gr:lam2S} we depict the mass of the
$\widetilde{\Xi}_c$ baryon and residues of the ground state $\Xi_c$,
$\widetilde{\Xi}_c$ and $\Xi'_c$ baryons as functions of these
parameters at fixed value of $\beta=-0.75$. From these figures, we see  weak dependence of the quantities under consideration on $M^2$ and $s_0$, satisfying the requirements of the method used.

To find the working region of $\beta = \tan\theta$, as 
examples, in Fig. 
\ref{gr:lam1PCosTeta} we present the dependence of the  $\widetilde{\Xi}_c$'s mass and residue
on  $\cos\theta$ at average values of $M^2$ and $s_0$. From this figure we see that the
results show relatively weak dependence  on the variations of $\cos\theta$
when it varies in the regions
\begin{equation}
-1\leq \cos\theta \leq -0.5, \, \, \,0.5\leq \cos\theta \leq 1.
\end{equation}
The errors coming from the variations of the results with respect to the variations of the auxiliary parameters remain within the limits allowed by the method used and they are included in final results. 

We depict the numerical results of the masses and residues of the
first angularly and radially excited $\Xi_c$ baryons as well as the residue of the ground state particle obtained using the above-presented working intervals for the auxiliary parameters in table
\ref{tab:NumResults}. Note that we get the same mass for the first angularly and radially excited $\Xi_c$ baryons. The errors in the presented results are due
to the uncertainties in the determination of the working regions for
the auxiliary parameters as well as the errors of other input
parameters. The values presented in table
\ref{tab:NumResults} will be used as inputs in next section.

\begin{table}[tbp]
\begin{tabular}{|c|c|c|c|}
\hline & $\Xi_c$ & $\widetilde{\Xi}_c$ & $\Xi'_c$\\ \hline \hline
$m~(\mathrm{MeV})$ &  & $2922\pm83$&$2922\pm83$ \\
\hline $\lambda  ~(\mathrm{GeV}^3)$ & $0.036\pm0.005$& $ 0.040\pm0.009$&$ 0.137\pm0.032$ \\
\hline
\end{tabular}%
\caption{The sum rule results for the masses and residues of the
first angularly and radially excited $\Xi_c$ baryon as well as
residue of the ground state. Note that the masses of the angularly and radially excited states are obtained to be exactly the same. } \label{tab:NumResults}
\end{table}

\section{ $ \widetilde{\Xi}_c$ and $\Xi'_c$ Transitions To $\Lambda_c K$ }

In this section we calculate the strong coupling constants
$g_{\widetilde{\Xi}_c\Lambda_cK}$ and $g_{\Xi'_c\Lambda_cK}$,
which are necessary to calculate widths of the decays $
\widetilde{\Xi}_c\rightarrow \Lambda_cK$ and $\Xi'_c\rightarrow
\Lambda_cK$. For this aim we introduce the correlation function
\begin{equation}
T (p,q)=i\int d^{4}xe^{ipx}\langle K(q)|\mathcal{T}\{J
_{\Lambda_{c}}(x) \bar{J }_{\Xi_c}(0)\}|0\rangle ,
\label{Eq:CorFunDecay}
\end{equation}%
where $J _{\Lambda_{c}}(x)$ is the interpolating current for the
$\Lambda_c$ baryon which can be obtained from  Eq.
(\ref{Eq:Current}) with  $q_1=u$ and $q_2=d$.

Firstly let consider the $\widetilde{\Xi}_c\rightarrow \Lambda_cK$
transition. Before calculations we note that the interpolating
current for $\Lambda_c$ interact with both positive and negative parity $\Lambda_c$
baryons. Taking into consideration this fact, inserting complete sets
of hadrons with the same quantum numbers as the interpolating
currents and isolating the ground states, we obtain

\begin{eqnarray}
T ^{\mathrm{Phys}}(p,q)&=&\frac{\langle 0|J _{\Lambda
_{c}}|\Lambda^+ _{c}(p,s)\rangle }{p^{2}-m_{\Lambda^+
_{c}}^{2}}\langle K(q)\Lambda^+ _{c}(p,s)|\Xi _{c}(p^{\prime
},s^{\prime })\rangle
\nonumber \\
&\times& \frac{\langle \Xi _{c}(p^{\prime },s^{\prime })|\bar{ J
}_{\Xi_c}|0\rangle }{p^{\prime 2}-m_{\Xi_c}^{2}}
 \nonumber
\\&+&
\frac{\langle 0|J _{\Lambda_{c}}|\Lambda^- _{c}(p,s)\rangle
}{p^{2}-m_{\Lambda^- _{c}}^{2}}\langle K(q)\Lambda^-
_{c}(p,s)|\widetilde{\Xi}
_{c}(p^{\prime },s^{\prime })\rangle  \nonumber \\
&\times& \frac{\langle \widetilde{\Xi} _{c}(p^{\prime },s^{\prime
})|\bar{ J }_{\widetilde{\Xi}_c}|0\rangle }{p^{\prime
2}-m_{\widetilde{\Xi}_c}^{2}}
\nonumber \\
&+&\frac{\langle 0|J _{\Lambda_{c}}|\Lambda^+ _{c}(p,s)\rangle
}{p^{2}-m_{\Lambda^+ _{c}}^{2}}\langle K(q)\Lambda^+
_{c}(p,s)|\widetilde{\Xi}
_{c}(p^{\prime },s^{\prime })\rangle  \nonumber \\
&\times& \frac{\langle \widetilde{\Xi} _{c}(p^{\prime },s^{\prime
})|\bar{ J }_{\widetilde{\Xi}_c}|0\rangle }{p^{\prime
2}-m_{\widetilde{\Xi}_c}^{2}}
\nonumber \\
&+& \frac{\langle 0|J _{\Lambda_{c}}|\Lambda^- _{c}(p,s)\rangle
}{p^{2}-m_{\Lambda^- _{c}}^{2}}\langle K(q)\Lambda^- _{c}(p,s)|\Xi
_{c}(p^{\prime },s^{\prime })\rangle  \nonumber \\
&\times& \frac{\langle \Xi _{c}(p^{\prime },s^{\prime })|\bar{ J
}_{\Xi_c}|0\rangle }{p^{\prime 2}-m_{\Xi_c}^{2}}
 +\ldots ,  \label{eq:SRDecay}
\end{eqnarray}
where $p^{\prime }=p+q,\ p$ and $q$ are the momenta of the $\Xi
_{c}$, $\Lambda _{c}$ baryons and $K$ meson, respectively.
$\Lambda^+_c$ and $\Lambda^-_c$ are the positive and negative
parity baryons in  spin-$1/2$ $\Lambda_c$ channel.
 In this expression $
m_{\Lambda _{c}}$ is the mass of the $\Lambda _{c}$ baryon. The dots in Eq.\ (%
\ref{eq:SRDecay}) stand for contributions of the higher resonances
and continuum states.

The matrix elements in Eq. (\ref{eq:SRDecay}) are parameterized as
\begin{eqnarray}
\langle 0|J _{\Lambda_c^+ }|\Lambda^+ _{c}(p,s)\rangle & =&\lambda
_{\Lambda^+ _{c}}u(p,s), \nonumber \\
\langle 0|J _{\Lambda_c^- }|\Lambda^-_{c}(p,s)\rangle &=&\lambda
_{\Lambda^-_{c}}\gamma_5 u(p,s),
\nonumber \\
\langle K(q)\Lambda^+ _{c}(p,s)|\Xi _{c}(p^{\prime },s^{\prime
})\rangle &=& g_{\Xi_c \Lambda^+_c K}\overline{u}(p,s)\gamma
_{5}u(p^{\prime },s^{\prime }),
\nonumber \\
\langle K(q)\Lambda^+ _{c}(p,s)|\widetilde{\Xi} _{c}(p^{\prime
},s^{\prime })\rangle &=& g_{\widetilde{\Xi}_c \Lambda^+_c
K}\overline{u}(p,s)u(p^{\prime },s^{\prime }),
\nonumber \\
\langle K(q)\Lambda^- _{c}(p,s)|\Xi _{c}(p^{\prime },s^{\prime
})\rangle &=& g_{\Xi_c \Lambda^-_c K}\overline{u}(p,s) u(p^{\prime
},s^{\prime }),
\nonumber \\
\langle K(q)\Lambda^- _{c}(p,s)|\widetilde{\Xi} _{c}(p^{\prime
},s^{\prime })\rangle &=&g_{\widetilde{\Xi}_c \Lambda^-_c
K}\overline{u}(p,s)\gamma_5 u(p^{\prime },s^{\prime }). \nonumber
\\  \label{eq14}
\end{eqnarray}
where $g_i$ are the strong coupling constants for corresponding
transitions.

Using the matrix elements given in Eq.(\ref{eq14}) and performing
summations over spins of $\Lambda_c$ and $\Xi_c$ baryons and
applying the double Borel transformations with respect $p^2$ and
$p^{\prime 2}$, for the physical side of the correlation function, we
get
\begin{eqnarray}
&&\mathcal{B}T ^{\mathrm{Phys}}(p,q)=g_{\Xi_c \Lambda^+_c
K}\lambda _{\Xi
_{c}}\lambda_{\Lambda_c^+}e^{-m_{\Xi_c}^{2}/M_{1}^{2}}e^{-m_{\Lambda
_{c}^+}^{2}/M_{2}^{2}}  \notag \\
&&\times \left( \slashed p+m_{\Lambda^+_{c}}\right)\gamma_5\left(
\slashed p+\slashed q+m_{\Xi_{c}}\right)   \notag \\
&&-g_{\widetilde{\Xi}_c \Lambda^-_c K}\lambda _{\widetilde{\Xi}
_{c}}\lambda_{\Lambda_c^-}e^{-m_{\widetilde{\Xi}_c}^{2}/M_{1}^{2}}
e^{-m_{\Lambda
_{c}^-}^{2}/M_{2}^{2}}  \notag \\
&&\times \gamma_5\left( \slashed
p+m_{\Lambda^-_{c}}\right)\gamma_5\left( \slashed p+\slashed
q+m_{\widetilde{\Xi}_{c}}\right)\gamma_5
\notag \\
&&-g_{\widetilde{\Xi}_c \Lambda^+_c K}\lambda _{\widetilde{\Xi}
_{c}}\lambda_{\Lambda_c^+}e^{-m_{\widetilde{\Xi}_c}^{2}/M_{1}^{2}}
e^{-m_{\Lambda
_{c}^+}^{2}/M_{2}^{2}}  \notag \\
&&\times \left( \slashed p+m_{\Lambda^-_{c}}\right)\left( \slashed
p+\slashed q+m_{\widetilde{\Xi}_{c}}\right)\gamma_5
\notag \\
&&+ g_{\Xi_c \Lambda^-_c K}\lambda _{\Xi
_{c}}\lambda_{\Lambda_c^-}e^{-m_{\Xi_c}^{2}/M_{1}^{2}}e^{-m_{\Lambda
_{c}^-}^{2}/M_{2}^{2}}  \notag \\
&&\times \gamma_5 \left( \slashed p+m_{\Lambda^-_{c}}\right)\left(
\slashed p+\slashed q+m_{\Xi_{c}}\right) , \label{eq:CFunc1/2}
\end{eqnarray}%
where  $M_{1}^{2}$ and $%
M_{2}^{2}$ are the Borel parameters.

From Eq. (\ref{eq:CFunc1/2}) follows that we have different
structures  which can be used to derive the sum rules for the
strong coupling constants for $\widetilde{\Xi}_{c}\rightarrow
\Lambda_{c}^{+}K$ channel. We have four couplings (see
Eq.\ref{eq:SRDecay}), and in order to determine the coupling
$g_{\widetilde{\Xi}_c^{\prime}\Lambda_cK}$ we need four equations.
Therefore we select the structures $\slashed q\slashed p
\gamma_5$, $\slashed p \gamma_5$, $\slashed q \gamma_5$ and
$\gamma_5$. Solving four algebraic equations for
$g_{\widetilde{\Xi}_{c}^{\prime}\Lambda^{+}K}$, we obtain
\begin{eqnarray}
g_{\widetilde{\Xi}_c \Lambda^+_c
K}&=&\frac{e^{\frac{m_{\widetilde{\Xi}_c}^2}{M_1^2}}e^{\frac{m_{\Lambda^+_c}^2}{M_2^2}}}
{\lambda_{\widetilde{\Xi}_c}\lambda_{\Lambda_c^+}(m_{\Lambda_c^+}+m_{\Lambda_c^-})
(m_{\Xi_c}+m_{\widetilde{\Xi}_c})} \nonumber \\
&\times& \left[T
_{1}^{\mathrm{QCD}}\left(m_K^2+m_{\Lambda_c^-}m_{\Xi_c}-m_{\Xi_c}
m_{\widetilde{\Xi}_c}\right)\right.
\nonumber \\
&+&\left.T
_{2}^{\mathrm{QCD}}\left(m_{\widetilde{\Xi}_c}-m_{\Lambda_c^-}
-m_{\Xi_c}\right) \right.
\nonumber \\
&+&\left. T
_{3}^{\mathrm{QCD}}\left(m_{\Lambda_c^-}-m_{\widetilde{\Xi}_c}
\right)-T _{4}^{\mathrm{QCD}}\right],\label{eq16}
\end{eqnarray}
where $T _{1}^{\mathrm{QCD}}$, $T
_{2}^{\mathrm{QCD}}$, $T _{3}^{\mathrm{QCD}}$
and $T _{4}^{\mathrm{QCD}}$ are the invariant
amplitudes corresponding to the structures $\slashed q\slashed
p\gamma _{5}$, $\slashed p\gamma _{5}$, $\slashed q\gamma _{5}$
and $\gamma _{5}$, respectively.

The general expressions obtained above contain two Borel
parameters $M_1^{2}$ and $M_2^{2}$. In our analysis we choose
\begin{equation}
M_{1}^{2}=M_{2}^{2}= 2M^{2},\,\,
M^2=\frac{M_1^{2}M_2^{2}}{M_1^{2}+M_2^{2}},
\end{equation}
which is traditionally justified by the fact that masses of the
involved heavy baryons $\Xi_c$ and $\Lambda_c$ are close to each
other.
The sum rules corresponding to the coupling constant defining the $\Xi'_c \rightarrow
\Lambda^+_cK$ transition can be easily obtained from Eq.(\ref{eq16}), by
replacing $m_{\widetilde{\Xi}_c}\rightarrow - m_{\Xi_c^{\prime}}$
and $\lambda_{\widetilde{\Xi}_c}\rightarrow
\lambda_{\Xi_c^{\prime}}$.

The QCD side of the correlation function for $\Pi
^{\mathrm{QCD}}(p,q)$ can be obtained by contracting out the quark fields using Wick's theorem and inserting into the obtained expression the relevant
quark propagators. For obtaining nonperturbative contributions in
light cone QCD sum rules, which are described in terms of  the $K$-meson
distribution amplitudes, one can use the  Fierz rearrangement
formula
\begin{equation*}
\overline{s}_{\alpha }^{a}u_{\beta }^{b}=\frac{1}{4}\Gamma _{\beta
\alpha }^{i}(\overline{s}^{a}\Gamma ^{i}u^{b}),
\end{equation*}%
where $\ \Gamma ^{i}=1,\ \gamma _{5},\ \gamma _{\mu },\ i\gamma
_{5}\gamma _{\mu },\ \sigma _{\mu \nu }/\sqrt{2}$ is the full set
of Dirac matrices. Sandwiched between the K-meson and vacuum
states, these terms as well as the ones generated by insertion of the
gluon field strength tensor $G_{\lambda \rho }(uv)$ \ from quark
propagators, give these distribution amplitudes (DAs) of various
quark-gluon contents in terms of wave functions with definite twists.
The DAs are main nonperturbative inputs of light cone QCD sum
rules. For  $K$-meson they  are derived in
\cite{Ball:2006wn,Belyaev:1994zk,Ball:2004ye}, which will be used in
our numerical analysis.  All these steps summarized above result
in a lengthy expression for the QCD side of correlation function. In
order not to overwhelm the study with overlong mathematical
expressions, we prefer not to present them here. Apart from
parameters in the distribution amplitudes, the sum rules for the
couplings depend also on numerical values of the $\Lambda_c^{+}$
baryon's mass and pole residue. In numerical calculations we
utilize
\begin{equation}
m_{\Lambda_c}=2286.46\pm 0.14 ~\mathrm{MeV} ,
\lambda_{\Lambda_c}=0.038 \pm 0.009 ~\mathrm{GeV}^3,
\end{equation}
where the value for the residue $ \lambda_{\Lambda_c} $ has been extracted from the corresponding mass sum rules in the present study and we use the mass of $ \Lambda_c $ state from PDG \cite{PDG}. 
 The working regions of the
Borel mass $M^2$, threshold $s_0$ and $\beta$ parameters for
calculations of the relevant strong couplings are chosen the same as the mass sum rules analyses.

Using the couplings $g_{\widetilde{\Xi}_c \Lambda^+_c K}$ and
$g_{\Xi'_c  \Lambda^+_c K}$ we can easily calculate the width of
$\widetilde{\Xi}_c \rightarrow \Lambda^+_c K$ and $\Xi'_c
\rightarrow \Lambda^+_c K$
decays. After some computations we obtain:%
\begin{eqnarray}
\Gamma \left( \widetilde{\Xi}_c \rightarrow \Lambda^+_c K\right)
&=&\frac{ g_{\widetilde{\Xi}_c \Lambda^+_c K}^{2}}{16\pi
m_{\widetilde{\Xi}_c}^{3}}\left[ (m_{\widetilde{\Xi}_c}
+m_{\Lambda^+_{c}})^{2}-m_{K}^{2}\right]  \notag \\
&&\times \lambda^{1/2}(m_{\widetilde{\Xi}_c}^2,m_{\Lambda^+
_{c}}^2,m_{K}^2),
\end{eqnarray}
and

\begin{eqnarray}
\Gamma \left( \Xi _{c}^{\prime }\rightarrow \Lambda _{c}^{+}K\right) &=&%
\frac{g_{\Xi ^{\prime }\Lambda^+_c K}^{2}}{16\pi m_{\Xi'_c}^{
3}}\left[ (m_{\Xi'_c}-m_{\Lambda^+ _{c}})^{2}-m_{K}^{2}\right]  \notag \\
&&\times \lambda^{1/2}(m_{\Xi'_c}^2,m_{\Lambda^+ _{c}}^2,m_{K}^2),
\end{eqnarray}
In expressions above the function $\lambda(x^2,y^2,z^2)$ is given
as:
\begin{equation*}
\lambda(x^2,y^2,z^2)=
x^{4}+y^{4}+z^{4}-2x^{2}y^{2}-2x^{2}z^{2}-2y^{2}z^{2}.
\end{equation*}

Numerical values obtained from our analyses for coupling constants and decay widths are
presented in table \ref{tab:NumResults1}. The obtained central value for
the decay width of the $\widetilde{\Xi}_c \rightarrow \Lambda^+_c
K$ case is in nice consistency with the central value of the experimental 
data, $[19.5\pm8.4(\mathrm{stat.})^{+5.4}_{-7.9}]~\mathrm{MeV}$
 \cite{Li:2017uvv}. 
 %Although the central value of our prediction for the width of $\Xi
%_{c}^{\prime }\rightarrow \Lambda _{c}^{+}K$ is considerably small
%compared to the corresponding experimental value, it has considerable overlap with the data considering the %uncertainties and can not be ruled out. 
In order to make a definite  conclusion about the nature of $\Xi_c(2930)$ state, more new  and refined experimental data with small errors are needed.

\begin{table}[tbp]
\begin{tabular}{|c|c|c|}
\hline &$ g$& $\Gamma (\mathrm{MeV})$\\
\hline \hline
$\widetilde{\Xi}_c \rightarrow \Lambda^+_c K$ &$0.66\pm0.11$  & $19.4\pm3.3$ \\
\hline $\Xi _{c}^{\prime }\rightarrow \Lambda _{c}^{+}K$ &
$7.15\pm2.21$& $ 13.6\pm2.3$
 \\
\hline
\end{tabular}%
\caption{The sum rule results for the strong coupling constants
and decay widths of the first angularly and radially excited
$\Xi_c$ baryon.} \label{tab:NumResults1}
\end{table}

\section{Summary and Concluding remarks}
We performed a QCD sum rule based analysis on the mass and width of the $\Xi_c(2930)$ considering it as first angularly/ radially excited charmed-strange baryon in $\Xi_c$ channel. We obtained the same mass for both the angularly and radially excited states,  and in excellent  agreement with the experimental value by Belle Collaboration, preventing us to assign any of these possibilities for the structure of this state. In next step, we considered the dominant decay of $\Xi_c(2930)$ to   $\Lambda_c K$ in both scenarios. 
The obtained central value for the width is nicely consistent with the central value of the experimental data of BELLE Collaboration,
when we consider $\Xi_c(2930)$ state as the angular excitation of the ground state charmed-strange baryon. This suggests the assignment of  a spin-parity
$J^P=\frac{1}{2}^{-}$ for this state. However, to make a final decision about the nature of  $\Xi_c(2930)$ state more new and refined data with small statistical and systematical errors are needed.\\

In Ref. \cite{Chen} and some other studies the orbitally excited single charmed baryons are classified into the  $\rho$ and $\lambda$ modes according to the quark model. When we compare  our results and assignment on $\Xi_c(2930)$ state with the results presented, for instance, in Ref. \cite{Chen}, we observe that  the $\Xi_c(2930)$ state is close to the $\lambda$ mode.
\appendix*

%%%%%%%%%%%%%%%%%%%%%%%%%%%%%%%%%%%%%%%%%%%%%%%%%%%%%%%%%%%%%%%%%

\section{ The QCD side of the correlation function}

\label{sec:App}
%%%%%%%%%%%%%%%%%%%%%%%%%%%%%%%%%%%%%%%%%%%%%%%%%%%%%%%%%%%%%%55
\renewcommand{\theequation}{\Alph{section}.\arabic{equation}}

In this appendix we present the explicit expressions of the  functions $ \mathcal{B}\Pi^{\mathrm{QCD}}_1(q) $ and $ \mathcal{B}\Pi^{\mathrm{QCD}}_2(q) $ used in mass sum rules:

\begin{widetext}

\begin{eqnarray}
\mathcal{B}\Pi^{\mathrm{QCD}}_1(q)&=&\int_{m_c^2}^{s_0}e^{-\frac{s}{M^2}}
\frac{1}{2^6\pi^2s^2}\left\{\frac{(5\beta^2+2\beta+5)}{32\pi^2}\left(8m_c^6s
-m_c^8-8m_c^2s^3+s^4+12m_c^4s^2\log\left[\frac{s}{m_c^2}\right]\right)
-\left(\langle
\bar{s}s\rangle+\langle \bar{d}d\rangle\right) \right.  \nonumber \\
 &\times& \left. \frac{m_c(m_c^2-s)^2}{3}(5\beta^2-4\beta-1)+
 \frac{\langle g_s^2
GG\rangle}{48\pi^2}
\left[3sm_c^2(1+\beta)^2-8m_c^4(1+\beta+\beta^2)+s^2(5+2\beta+5\beta^2)
\right]
\right. \nonumber \\
 &+&\left. \frac{m_0^2 \left(\langle
\bar{s}s\rangle+\langle \bar{d}d\rangle\right)}{12}m_c(\beta-1)
\left[m_c^2(11\beta+7)-6s(\beta+1)\right]-\frac{\langle g_s^2
GG\rangle \left(\langle \bar{s}s\rangle+\langle
\bar{d}d\rangle\right)}{6}m_c(\beta^2-1) \right\}
\nonumber \\
 &+&e^{-\frac{m_c^2}{M^2}}\left\{\frac{\langle
\bar{s}s\rangle\langle
\bar{d}d\rangle}{72}\left(11\beta^2+2\beta-13\right)+\frac{\langle
g_s^2 GG\rangle \left(\langle \bar{s}s\rangle+\langle
\bar{d}d\rangle\right)}{864\pi^2m_c}\left(\beta^2+\beta-2\right)-
\frac{\langle g_s^2 GG\rangle^2}{221184\pi^4M^2}
\right. \nonumber \\
 &\times&\left. \left(13\beta^2+10\beta+13\right)
 -\frac{m_0^2\langle \bar{s}s\rangle\langle
\bar{d}d\rangle}{288M^4}(\beta-1)\left[m_c^2(26+22\beta)+M^2(25+23\beta)\right]
-\frac{\langle g_s^2 GG\rangle m_0^2\left(\langle
\bar{s}s\rangle+\langle
\bar{d}d\rangle\right)}{55296\pi^2m_cM^4}
\right. \nonumber \\
 &\times&\left.(\beta-1)\left[m_c^2(11\beta+31)-2M^2(5\beta+1)\right]
 -\frac{\langle g_s^2 GG\rangle\langle \bar{s}s\rangle\langle
\bar{d}d\rangle}{5184M^6}m_c^2\left(11\beta^2+2\beta-13\right)
+\frac{\langle g_s^2 GG\rangle m_0^2\langle \bar{s}s\rangle\langle
\bar{d}d\rangle}{10368M^{10}}
\right. \nonumber \\
 &\times&\left.
m_c^2(m_c^2-2M^2)\left(11\beta^2+2\beta-13\right) \right. \bigg\},
\end{eqnarray}

\begin{eqnarray}
\mathcal{B}\Pi^{\mathrm{QCD}}_2(q)&=&\int_{m_c^2}^{s_0}e^{-\frac{s}{M^2}}
\frac{1}{3\cdot2^6\pi^2s}\left\{\frac{m_c\left(11\beta^2+2\beta-13\right)}
{2^3\pi^2}
\left[s^3+9m_c^2s^2-9m_c^4s-m_c^6-6m_c^2s\left(s+m_c^2\right)
\log[\frac{s}{m_c^2}]\right] \right.
\nonumber \\
&-&  \left(\langle \bar{s}s\rangle+\langle \bar{d}d\rangle\}
\right)(5\beta^2-4\beta-1)(m_c^2-s)^2 +\frac{\langle g_s^2
GG\rangle (\beta-1)}{3\cdot2^4 \pi^2 m_c}
\left[(s-m_c^2)\left(s(11\beta+13)+m_c^2(67\beta+53)\right)\right.
\nonumber \\
&-& \left.\left. 3m_c^2s
(13\beta+11)\log[\frac{s}{m_c^2}]\right]+\frac{m_0^2\left(\langle
\bar{s}s\rangle+\langle \bar{d}d\rangle\right)}{2^2}(\beta
-1)\left(6s(\beta+1)-m_c^2(\beta+5)\right) \right\}
\nonumber \\
&+&e^{-\frac{m_c^2}{M^2}}\left\{\frac{1}{24}\langle
\bar{s}s\rangle\langle \bar{d}d\rangle
m_c(5\beta^2+2\beta+5)+\frac{\langle g_s^2 GG\rangle\left(\langle
\bar{s}s\rangle+\langle
\bar{d}d\rangle\right)}{3^3\cdot2^7\pi^2}(\beta^2-8\beta+7)-\frac{\langle
g_s^2 GG\rangle^2}{3^3\cdot2^{13}\pi^4M^2m_c} \right.
\nonumber \\
&\times& \left.
(m_c^2-2M^2)(13\beta^2-2\beta-11)+\frac{m_0^2\langle
\bar{s}s\rangle\langle \bar{d}d\rangle
m_c}{3^2\cdot2^4M^4}\left[M^2(\beta-1)^2-3m_c^2(5\beta^2+2\beta+5)\right]
\right.
\nonumber \\
&+& \left. \frac{\langle g_s^2 GG\rangle m_0^2(\langle
\bar{s}s\rangle+\langle \bar{d}d\rangle)m_c^2}{3^3\cdot2^{11}\pi^2
M^4}(\beta^2+28\beta-29)-\frac{\langle g_s^2 GG\rangle\langle
\bar{s}s\rangle\langle \bar{d}d\rangle m_c}{3^3\cdot2^6
M^6}(m_c^2-3M^2)(5\beta^2+2\beta+5) \right.
\nonumber \\
&+& \left. \frac{\langle g_s^2 GG\rangle m_0^2\langle
\bar{s}s\rangle\langle \bar{d}d\rangle m_c}{3^3\cdot2^7
M^{10}}\left(m_c^4-6m_c^2M^2+6M^4\right)(5\beta^2+2\beta+5)
 \right\}.
\end{eqnarray}

In calculations we set $m_u=m_d=0$ but $m_s\neq0$. To shorten
the above expressions, the   terms proportional to $m_s$ have  not been presented, but their
contributions have been taken into account when performing the numerical
analysis.

\section*{ACKNOWLEDGEMENTS}
 K.A. thanks  Dogu\c{s} University for partial financial support provided under contract BAP 2015-16-D1-B04.
 H. S. also would like to thank Kocaeli University for the partial financial support through the grant BAP 2018/070.
\end{widetext}
%

%%%%%%%%%%%%%%%%%%%%%%%%%%%%%%%%%%%%%%%%%%%%%%%%%%%%%%%%%%%%%%%%%%%%%%%%%%%%%%%

\end{document}